# Low-Frequency Noise in Low-Dimensional van der Waals Materials

Alexander A. Balandin
*Nano-Device Laboratory*
*Department of Electrical and Computer Engineering*
*University of California – Riverside*
Riverside, California 92521 USA
E-mail: balandin@ece.ucr.edu

Sergey Rumyantsev
*Center for Terahertz Research and Applications*
*Institute of High Pressure Physics*
*Polish Academy of Sciences*
Warsaw, 01-142 Poland
E-mail: roumis4@gmail.com

*Abstract* – The emergence of graphene and two-dimensional van der Walls materials renewed interest to investigation of the low-frequency noise in the low-dimensional systems. The layered van der Waals materials offers unique opportunities for studying the low-frequency noise owing to the properties controlled by the thickness of these materials, and tunable carrier concentration. In this review, we describe unusual low-frequency noise phenomena in quasi-2D and quasi-1D van der Waals materials. We also demonstrate that the low-frequency noise spectroscopy is a powerful tool for investigation of the electron transport and charge-density-wave phase transitions in this class of materials.

## I. INTRODUCTION

The emergence of graphene and two-dimensional (2D) *van der Walls* materials renewed interest to investigation of the low-frequency noise in the low-dimensional systems [1-3]. This type of materials offers unique opportunities for studying the low-frequency noise phenomena owing to their properties, controlled by the film thickness, and their widely tunable charge carrier concentration. Practical applications depend on the ability to understand and reduce the noise in this new type of materials. From the other side, these materials offer new opportunities for addressing the fundamental problems of noise and fluctuations.

## II. NOISE IN GRAPHENE

The first material of this class – graphene – revealed a number of interesting properties in the context of $1/f$ noise owing to its 2D nature, unusual linear energy dispersion for electrons and holes, zero-energy bandgap, specific scattering mechanisms and metallic-type conductance [1-15]. From one side, graphene is an ultimate surface where conduction electrons are exposed to the traps, e.g. charged impurities in a substrate or on its top surface, which can result in strong carrier-number fluctuations. From the other side, graphene can be considered a zero-bandgap metal, where mobility fluctuations resulting from the charged scattering centers in the substrate or surface can also make a strong contribution to $1/f$ noise. The ability to change the thickness of few-layer graphene conductors by one atomic layer at a time opened up opportunities for examining surface and volume contributions to $1/f$ noise directly [14-15]. Independent studies have shown that noise in graphene reveals an unusual gate-bias dependence [5-13]. In the vicinity of the Dirac point, the noise amplitude follows a V-shape dependence, attaining its minimum at the Dirac point where the resistance is at its maximum [10]. The unusual gate dependence of the noise amplitude in graphene supported the conclusion that $1/f$ noise in graphene devices does not follow the conventional McWhorter model used for complementary metal-oxide-semiconductor (CMOS) devices and other metal-oxide-semiconductor field-effect transistors (MOSFETs) (see Figure 1).

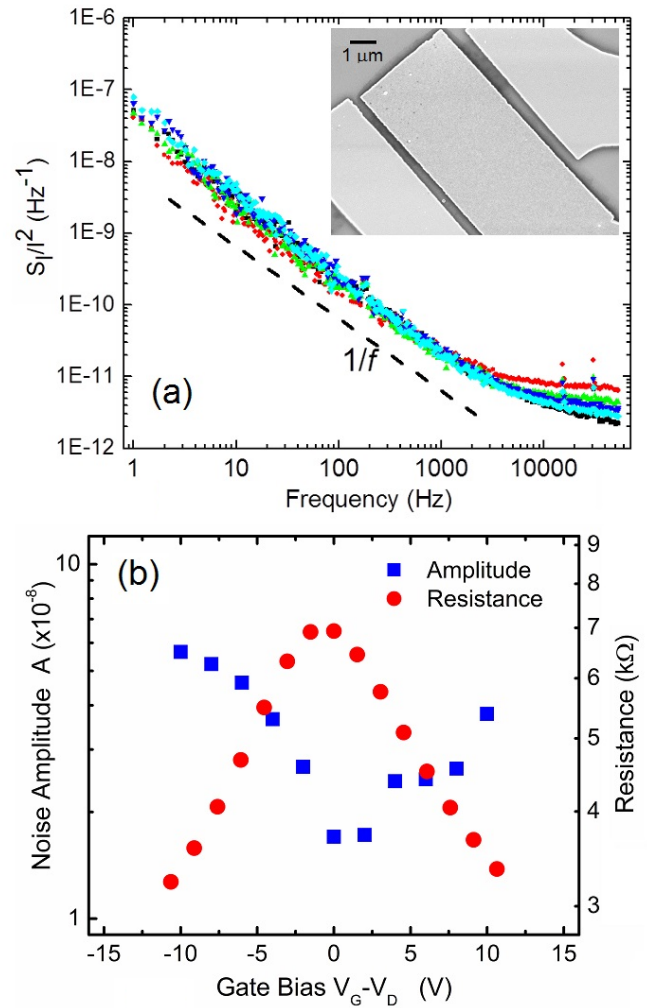

Fig. 1 Low-frequency noise in graphene. (a) Typical $1/f$ noise spectra in a graphene device. The inset shows a scanning electron microscopy image of a representative graphene device. (b) Unusual V-shape bias dependence of $1/f$ noise in graphene. The figures are reproduced from Ref. [1] with permission from the Nature Publishing Group.

## III. NOISE IN 2D VAN DER WAALS MATERIALS

Recently, the exotic phenomena, such as charge density waves (CDW) in 2D van der Waals materials, attracted interest in the context of the low-frequency noise research [16-18]. The CDW phase is a macroscopic quantum state consisting of a periodic modulation of the electronic charge density accompanied by a periodic distortion of the atomic lattice [19]. One of the most interesting 2D CDW materials is $1T$-$TaS_2$ [17, 20-23]. As the temperature increases above 180 K, the commensurate CDW phase in this material breaks up into a nearly-commensurate CDW phase that consists of ordered commensurate CDW regions separated by domain walls. This transition is revealed as an abrupt change in the resistance with a large hysteresis window in the resistance profile at 200 K. As the temperature increases to 350 K, the nearly commensurate phase melts into an incommensurate phase, in which the CDW wave vector is no longer commensurate with the lattice. This transition is accompanied by a smaller hysteresis window in the resistivity. Only at high temperatures of 500 K – 600 K the incommensurate CDW phase melts into the normal metallic phase of $1T$-$TaS_2$.

temperature stimuli [17]. This study revealed that noise in $1T$-$TaS_2$ has several pronounced maxima at the bias voltages, which correspond to the onset of CDW sliding and to the phase transitions (see Figure 2). The noise spectral density was more sensitive to the phase transitions and changes in the electron transport than the current-voltage (I-V) characteristics. We observed the unusual Lorentzian features and exceptionally strong noise dependence on the electric bias and temperature, leading to the conclusion that electronic noise in 2D CDW systems has a unique physical origin different from known fundamental noise types [17]. It was established that the low-frequency noise spectroscopy can serve as useful tool for understanding electronic transport phenomena in 2D CDW materials characterized by coexistence of different phases and strong pinning. The technique has been also used for the vertical $1T$-$TaS_2$ CDW devices. It was found that the low-frequency noise spectral density revealed strong peaks, below the temperature of the commensurate CDW to nearly-commensurate CDW transition, possibly indicating the presence of the strongly debated *hidden phase* states [18].

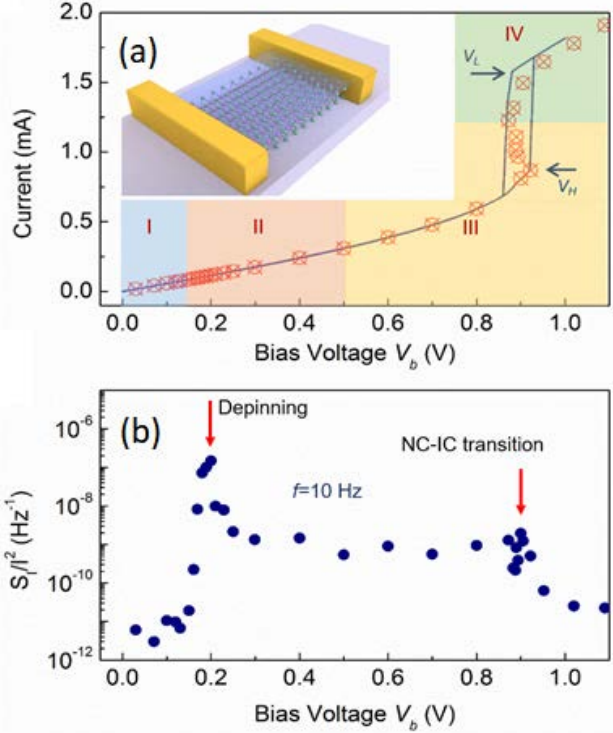

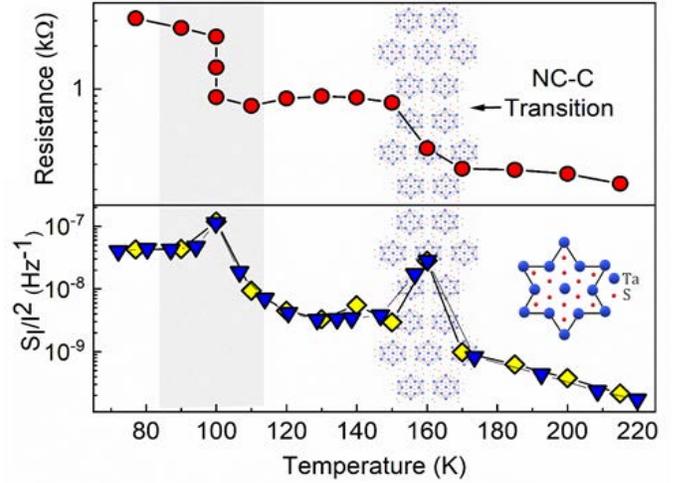

Fig. 2 Low frequency noise in 2D charge-density-wave materials. (a) Current-voltage characteristics of thin-film $1T$-$TaS_2$ device at room temperature. The inset shows the device schematic. The abrupt CDW phase transitions are seen at voltages $V_H$ (up scan) and $V_L$ (down scan). The red circles indicated the biasing points for the low-frequency noise measurements. (b) The normalized noise spectral density, $S_I/I^2$ at 10 Hz as a function of the bias voltage, $V_B$. Two pronounced local maxima correspond to the de-pinning of the charge density wave and the phase transition between two CDW phases. The figures are adapted from Ref. [17].

We investigated the low-frequency noise in $1T$-$TaS_2$ thin films as they were driven from the nearly commensurate (NC) to incommensurate (IC) CDW phases by voltage and

Fig. 3 Electrical resistance (upper panel) and normalized current noise spectral density (lower panel) as the functions of temperature. The noise data were measured at the bias of $V_{SD}=13$ mV and frequency $f=10$ Hz. The decrease in resistance at $T_C=160$ K corresponds to the well-known commensurate to nearly-commensurate CDW transition. A diagram, depicting the reconstruction of Ta atoms into hexagonal clusters, is shown to illustrate the phase transition. A distinctive noise peak is observed at the same temperature $T_C$. Below the commensurate CDW – nearly-commensurate CDW phase transition temperature, one can see another step in the resistance with the corresponding peak in the noise spectral density. The figure is adapted from Ref. [18]

The low-frequency electronic noise spectroscopy has been instrumental in monitoring the electric-field induced transition from the incommensurate CDW phase to the normal metal phase. The noise spectral density, $S_I/I^2$, exhibits sharp increases at the phase transition points, which correspond to the step-like changes in resistivity. Assignment of the phases was consistent with low-field resistivity measurements over the temperature range from 77 K to 600 K. The transition to the metallic phase of $1T$-$TaS_2$ is not accompanied by a large change in the resistivity. However, noise spectrum undergoes substantial changes, which allow to accurately determine the transition (see Figure 4).

## IV. NOISE IN 1D VAN DER WAALS MATERIALS

The concept of van der Waals materials has been extended to one-dimensional (1D) systems. In contrast to the layered quasi-2D crystals, quasi-1D materials, such as $MX_3$ (where M = transition metals; X = Se and Te) consist of the atomic threads, which are weakly bound in bundles by van der Waals forces. As a consequence, the exfoliation of the $MX_3$ crystals results not in 2D layers but rather in quasi-1D nanowires [24-28]. It has been shown that some of quasi-1D van der Waals materials reveal exceptionally high electrical current densities [24, 27]. These materials are very interesting from the low-frequency noise prospective as well (see Figure 5).

Specifically, we found that quasi-1D $TaSe_3$ and $ZrTe_3$ van der Waals nanowires, which possess exceptionally high current densities, have rather low levels of the low-frequency noise as compared to graphene [25, 28]. In $ZrT_3$ nanowires, the noise reveals the $1/f$ behavior near room temperature but becomes dominated by the Lorentzian bulges at low temperature. Unexpectedly, the corner frequency of the Lorentzian peaks shows a strong sensitivity to the applied source–drain bias. The dependence on electric field can be explained by the Frenkel–Poole effect only in the scenario where the bias voltage drop happens predominantly on the defects, which block the quasi-1D conduction channels [28]. In $TaSe_3$, $1/f$ noise becomes the $1/f^2$ type as temperature increases to ~400 K, suggesting the onset of electromigration. Using Dutta−Horn model, we determined that the noise activation energy for quasi-1D $TaSe_3$ nanowires is ~1.0 eV, comparable to that for Cu and Al interconnects. Our results suggest that quasi-1D van der Waals metallic nanowires have potential for applications in the *ultimately downscaled* local interconnects [25]. While this review emphasized the use of low-frequency noise as a spectroscopy tool for understanding the electron transport phenomena, the obtained results are also important for practical applications of 2D and 1D van der Waals materials in sensors or information processing [29-35].

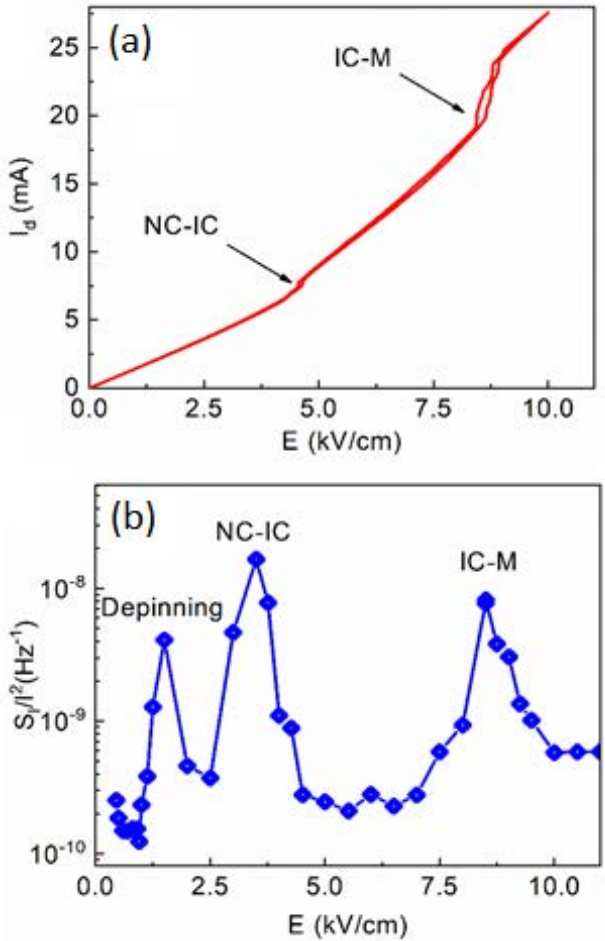

Fig. 4 Noise spectroscopy of CDW – metallic phase transition. (a) Current-voltage characteristics showing two steps in current at the electric field of 4.5 kV/cm and 9 kV/cm corresponding to the nearly-commensurate to incommensurate CDW phase transition and incommensurate to metal transition, respectively. (b) Normalized spectral density as a function of the electric field measured at $f$=10 Hz. The peaks in the noise spectral density at 4.5 kV/cm and 9 kV/cm are in excellent agreement with the resistance steps in (a). Note that the noise spectrum is much more sensitive to the phase transitions than the I-V characteristics.

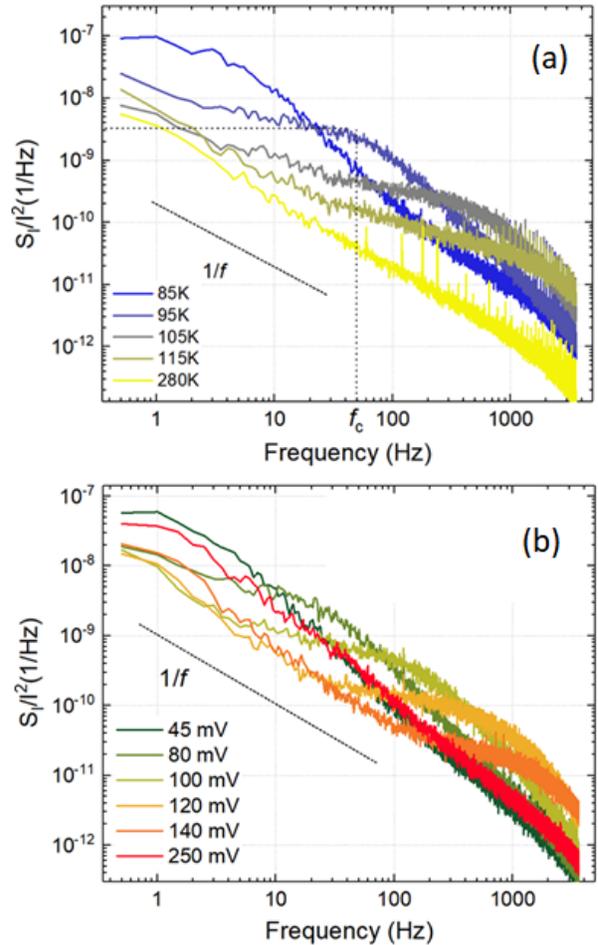

Fig. 5 Noise in 1D van der Waals materials. (a) Normalized noise spectral density, $S_I/I^2$, as a function of frequency of quasi-1D $ZrTe_3$ nanoribbon at temperatures from 85 K to 280 K. (a) Normalized noise spectral density, $S_I/I^2$, as a function of frequency of quasi-1D $ZrTe_3$ nanoribbon at the bias voltage ranging from 45 mV to 250 mV. Adapted from Ref. [28].

## V. CONCLUSIONS

We described our recent results pertinent to the unusual noise phenomena in quasi-2D and quasi-1D van der Waals materials. We also discussed the prospects of the low-frequency noise spectroscopy which is a powerful tool for investigation of the electron transport and charge-density-wave phase transitions in this class of materials.

## ACKNOWLEDGEMENTS

Device fabrication and testing were supported, in part, by the Semiconductor Research Corporation (SRC) contract 2018-NM-2796, the National Science Foundation (NSF) through the Emerging Frontiers of Research Initiative (EFRI) 2-DARE project NSF EFRI-1433395), and as part of the Spins and Heat in Nanoscale Electronic Systems (SHINES), and Energy Frontier Research Center funded by the U.S. Department of Energy, Office of Science, Basic Energy Sciences (BES) under Award No. SC0012670. S. R. acknowledges support from Center for Terahertz Research and Applications project carried out within the International Research Agendas program of the Foundation for Polish Science co-financed by the European Union under the European Regional Development Fund.

## REFERENCES

[1] A. A. Balandin, Low-frequency 1/$f$ noise in graphene devices, Nature Nano, 8, 549–555 (2013).

[2] M. A. Stolyarov, G. Liu, S. L. Rumyantsev, M. Shur, and A. A. Balandin, Suppression of 1/$f$ noise in near-ballistic h-BN-graphene-h-BN HFETs, Appl. Phys. Lett., 107, 23106 (2015).

[3] S. L. Rumyantsev, C. Jiang, R. Samnakay, M. S. Shur, and A. A. Balandin, 1/$f$ noise characteristics of $MoS_2$ thin-film transistors, IEEE Electron Device Letters, 36, 517 (2015).

[4] Z. Chen, Y. M. Lin, M. J. Rooks, and P. Avouris, Graphene nano-ribbon electronics, Physica E, 40, 228–232 (2007).

[5] Y. M. Lin and P. Avouris, Strong suppression of electrical noise in bilayer graphene Nanodevices, Nano Lett. 8, 2119–2125 (2008).

[6] A. N. Pal and A. Ghosh, Resistance noise in electrically biased bilayer graphene, Phys. Rev. Lett. 102, 126805 (2009).

[7] Q. Shao, et al. Flicker noise in bilayer graphene transistors, IEEE Electron. Dev. Lett. 30, 288–290 (2009).

[8] G. Xu, et al. Low-noise submicron channel graphene nanoribbons, Appl. Phys. Lett. 97, 073107 (2010).

[9] G. Xu, et al. Effect of spatial charge inhomogeneity on 1/$f$ noise behavior in graphene, Nano Lett. 10, 3312–3317 (2010).

[10] S. Rumyantsev, G. Liu, W. Stillman, M. Shur, and A. A. Balandin, Electrical and noise characteristics of graphene field-effect transistors: ambient effects, noise sources and physical mechanisms, J. Phys. Condensed Matter, 22, 395302 (2010).

[11] G. Liu, W. Stillman, S. Rumyantsev, M. Shur, and A. A. Balandin, Low-frequency electronic noise in graphene transistors: comparison with carbon nanotubes, Int. J. High Speed Electronic Syst., 20, 161–170 (2011).

[12] G. Liu, S. Rumyantsev, M. Shur, and A. A. Balandin, Graphene thickness-graded transistors with reduced electronic noise, Appl. Phys. Lett. 100, 033103 (2012).

[13] A. A. Kaverzin, A. S. Mayorov, A. Shytov, and D. W. Horsell, Impurities as a source of 1/$f$ noise in graphene, Phys. Rev. B 85, 075435 (2012).

[14] G. Liu, S. Rumyantsev, M. Shur, and A. A. Balandin, Origin of 1/$f$ noise in graphene multilayers: Surface vs. volume, Appl. Phys. Lett. 102, 093111 (2013).

[15] M. Hossain, S. L. Rumiantsev, M. Shur, and A. A. Balandin, Reduction of 1/f noise in graphene after electron-beam irradiation. Appl. Phys. Lett. 102, 153512 (2013).

[16] G. Liu, B. Debnath, T. R. Pope, T. T. Salguero, R. K. Lake, and A. A. Balandin, A charge-density-wave oscillator based on an integrated tantalum disulfide–boron nitride–graphene device operating at room temperature, Nature Nano, 11, 845 (2016).

[17] G. Liu, S. Rumyantsev, M. A. Bloodgood, T. T. Salguero, and A. A. Balandin, Low-frequency current fluctuations and sliding of the charge density waves in two-dimensional materials, Nano Lett., 18, 3630 (2018).

[18] R. Salgado, A. Mohammadzadeh, F. Kargar, A. Geremew, C.-Y. Huang, M. A. Bloodgood, S. Rumyantsev, T. T. Salguero, and A. A. Balandin, Low-frequency noise spectroscopy of charge-density-wave phase transitions in vertical quasi-2D 1T-$TaS_2$ devices, Appl. Phys. Express, 18, 037001 (2019).

[19] G. Grüner, The dynamics of charge-density waves, Rev. Mod. Phys., 60, 1129−1181 (1988).

[20] Y. I. Joe, et al., Emergence of charge density wave domain walls above the superconducting dome in 1T-$TiSe_2$, Nature Phys., 10, 421−425 (2014).

[21] M. J. Hollander, et al., Electrically driven reversible insulator-metal phase transition in 1T-$TaS_2$, Nano Lett., 15, 1861−1866 (2015).

[22] G. Liu, et al., Total-ionizing-dose effects on threshold switching in 1T-$TaS_2$ charge density wave devices, IEEE Electron Device Lett., 38, 1724–1727 (2017).

[23] A. Geremew, et al., Proton-irradiation-immune electronics implemented with two-dimensional charge-density-wave devices, Nanoscale, 11, 8380 (2019).

[24] M. A. Stolyarov, et al., Breakdown current density in h-BN-capped quasi-1D $TaSe_3$ metallic nanowires: prospects of interconnect applications, Nanoscale, 8, 15774–15782 (2016).

[25] G. Liu, S. Rumyantsev, M. A. Bloodgood, T. T. Salguero, M. Shur, and A. A. Balandin, Low-frequency electronic noise in quasi-1D $TaSe_3$ van der Waals nanowires, Nano Lett., 17, 377 (2017).

[26] M. A. Bloodgood, et al., Monoclinic structures of niobium trisulfide, APL Mater., 6, 026602 (2018).

[27] A. Geremew, et al., Current carrying capacity of quasi-1D $ZrTe_3$ van der Waals nanoribbons, IEEE Electron Device Lett., 39, 735–738 (2018).

[28] A. K. Geremew, S. Rumyantsev, M. A. Bloodgood, T. T. Salguero, and A. A. Balandin, Unique features of the generation–recombination noise in quasi-one-dimensional van der Waals nanoribbons, Nanoscale, 10, 19749 (2018).

[29] X. Yang, et al., Triple-mode single-transistor graphene amplifier and its applications, ACS Nano, 4, 5532–5538 (2010).

[30] X. Yang, et al., Graphene ambipolar multiplier phase detector, IEEE Electron Device Lett., 32, 1328–1330 (2011).

[31] S. Rumyantsev, G. Liu, M. S. Shur, R. A. Potyrailo, and A. A. Balandin, Selective gas sensing with a single pristine graphene transistor," Nano Lett., 12, 2294–2298 (2012).

[32] S. Rumyantsev, G. Liu, R. A. Potyrailo, A. A. Balandin, and M. S. Shur, Selective sensing of individual gases using graphene devices, IEEE Sens. J., 13, 2818–2822 (2013).

[33] G. Liu, S. Ahsan, A. G. Khitun, R. K. Lake, and A. A. Balandin, Graphene-based non-Boolean logic circuits, J. Appl. Phys., 114, 154310 (2013).

[34] A. Khitun, et al., Two-dimensional oscillatory neural network based on room-temperature charge-density-wave devices, IEEE Trans. Nanotech., 16, 860–867 (2017).

[35] A. G. Khitun, et al., Transistor-less logic circuits implemented with 2-D charge density wave devices, IEEE Electron Device Lett., 39, 1449–1452 (2018).

[36] A. K. Geremew, et al., Bias-voltage driven switching of the charge-density-wave and normal metallic phases in 1T-$TaS_2$ thin-film devices, ACS Nano, 13, 7231−7240 (2019).

[37] T. A. Empante, et al., Low resistivity and high breakdown current density of 10 nm diameter van der Waals TaSe$_3$ nanowires by chemical vapor deposition, Nano Lett., 19, 4355–4361 (2019).